\documentclass[amsmath,amssymb,aps, prl, twocolumn, superscriptaddress]{revtex4-2}
\usepackage{graphicx}% Include figure files
\usepackage{dcolumn}% Align table columns on decimal point
\usepackage{bm}% bold math
\usepackage{gensymb} % pour le symbole degree

\usepackage{xcolor}
\begin{document}

\preprint{APS/123-QED}

\title{Phase transitions of $\text{Fe}_2\text{O}_3$ under laser shock compression}
%\thanks{A footnote to the article title} 

\author{A. Amouretti}
\email{AmourettiA@eie.eng.osaka-u.ac.jp}
%\altaffiliation[Also at ]{Graduate School of Engineering, Osaka University, Suita, Osaka 565-0871, Japan}
\affiliation{IMPMC, Sorbonne Université, UMR CNRS 7590, MNHN, 75005 Paris, France}
\affiliation{Graduate School of Engineering, Osaka University, Suita, Osaka 565-0871, Japan}

\author{C. Crépisson}
\email{celine.crepisson@physics.ox.ac.uk}
\affiliation{Department of Physics, Clarendon Laboratory, University of Oxford, Parks Road, Oxford OX1 3PU, UK}

\author{S. Azadi}%
\affiliation{Department of Physics, Clarendon Laboratory, University of Oxford, Parks Road, Oxford OX1 3PU, UK}

\author{D. Cabaret}%
\affiliation{IMPMC, Sorbonne Université, UMR CNRS 7590, MNHN, 75005 Paris, France}
\author{T. Campbell}
\affiliation{Department of Physics, Clarendon Laboratory, University of Oxford, Parks Road, Oxford OX1 3PU, UK}
\author{D. A. Chin}
\affiliation{University of Rochester Laboratory for Laser Energetics, Rochester, NY, USA}
\author{B. Colin}
\affiliation{University of Rochester Laboratory for Laser Energetics, Rochester, NY, USA}
\author{G. R. Collins}
\affiliation{University of Rochester Laboratory for Laser Energetics, Rochester, NY, USA}
\author{L. Crandall}
\affiliation{University of Rochester Laboratory for Laser Energetics, Rochester, NY, USA}

\author{G. Fiquet}%
\affiliation{IMPMC, Sorbonne Université, UMR CNRS 7590, MNHN, 75005 Paris, France}
\author{A. Forte}%
\affiliation{Department of Physics, Clarendon Laboratory, University of Oxford, Parks Road, Oxford OX1 3PU, UK}

\author{T. Gawne}%
\affiliation{Department of Physics, Clarendon Laboratory, University of Oxford, Parks Road, Oxford OX1 3PU, UK}
\author{F. Guyot}%
\affiliation{IMPMC, Sorbonne Université, UMR CNRS 7590, MNHN, 75005 Paris, France}

\author{P. Heighway}%
\affiliation{Department of Physics, Clarendon Laboratory, University of Oxford, Parks Road, Oxford OX1 3PU, UK}

\author{H. Lee}
\affiliation{SLAC National Accelerator Laboratory, 2575 Sand Hill Rd, Menlo Park, CA 94025, USA}

\author{D. McGonegle}%
\affiliation{Atomic Weapons Establishment, Reading, UK}

\author{B. Nagler}
\affiliation{SLAC National Accelerator Laboratory, 2575 Sand Hill Rd, Menlo Park, CA 94025, USA}

\author{J. Pintor}%
\affiliation{IMPMC, Sorbonne Université, UMR CNRS 7590, MNHN, 75005 Paris, France}
\author{D. Polsin}
\affiliation{University of Rochester Laboratory for Laser Energetics, Rochester, NY, USA}

\author{G. Rousse}%
\affiliation{CSE Lab, UMR 8260, Collège de France, 75231 Paris Cedex 05, France}
\affiliation{Sorbonne Université, 4 place Jussieu, 75005 Paris, France}

\author{Y. Shi}
\affiliation{Department of Physics, Clarendon Laboratory, University of Oxford, Parks Road, Oxford OX1 3PU, UK}
\author{E. Smith}
\affiliation{University of Rochester Laboratory for Laser Energetics, Rochester, NY, USA}

\author{J. S. Wark}%
\affiliation{Department of Physics, Clarendon Laboratory, University of Oxford, Parks Road, Oxford OX1 3PU, UK}

\author{S. M. Vinko}%
\affiliation{Department of Physics, Clarendon Laboratory, University of Oxford, Parks Road, Oxford OX1 3PU, UK}
\affiliation{Central Laser Facility, STFC Rutherford Appleton Laboratory, Didcot OX11 0QX, UK}

\author{M. Harmand}%
\affiliation{IMPMC, Sorbonne Université, UMR CNRS 7590, MNHN, 75005 Paris, France}
\affiliation{PIMM, Arts et Metiers Institute of Technology, CNRS, Cnam, HESAM University, 151 boulevard de l’Hopital, 75013 Paris (France)}

\date{\today}

\begin{abstract}
We present {\it in-situ} x-ray diffraction and velocity measurements of $\text{Fe}_2\text{O}_3$ under laser shock compression at pressures between 38-116~GPa. None of the phases reported by static compression studies were observed. Instead, we observed an isostructural phase transition from $\alpha$-$\text{Fe}_2\text{O}_3$ to a new $\alpha'$-$\text{Fe}_2\text{O}_3$ phase at a pressure of 50-62 GPa. The $\alpha'$-$\text{Fe}_2\text{O}_3$ phase differs from $\alpha$-$\text{Fe}_2\text{O}_3$ by an 11\% volume drop and a different unit cell compressibility. We further observed a two-wave structure in the velocity profile, which can be related to an intermediate regime where both $\alpha$ and $\alpha'$ phases coexist. Density functional theory calculations with a Hubbard parameter indicate that the observed unit cell volume drop can be associated with a spin transition following a magnetic collapse.
\end{abstract}

\maketitle

The phase diagrams of iron oxides are notoriously rich with a variety of electronic and structural transitions triggered by pressure or temperature. Due to its relevance to geophysical studies, $\text{Fe}_2\text{O}_3$ has been extensively studied under static compression using Laser Heated Diamond Anvil Cells (LH-DAC), up to approximately 113 GPa and 2800 K \cite{bykova_structural_2016, ono_situ_2005, ono_high-pressure_2004}. A series of phase transitions were observed, as well as a possible breakdown of $\text{Fe}_2\text{O}_3$ at high temperature into Fe$_{25}$O$_{32}$ and Fe$_5$O$_7$ \cite{bykova_structural_2016}.  Moreover, a Mott transition and a high-spin to low-spin transition have been evidenced at $\sim$50 GPa, although it remains unclear if structural transitions are triggered by the electronic transition or vice versa \cite{pasternak_breakdown_1999, badro_nature_2002, sanson_local_2016, greenberg_pressure-induced_2018}.
The $\alpha$-$\text{Fe}_2\text{O}_3$ phase ($R\bar{3}c$) is stable up to 40 GPa. For this phase, a continuous decrease in the c/a ratio is observed with increasing pressure  
\cite{bykova_novel_2013, liu_static_2003, schouwink_high-pressure_2011, rozenberg_high-pressure_2002, finger_crystal_1980}. From 40 GPa to 47 GPa, the $\iota$-$\text{Fe}_2\text{O}_3$ phase (Rh$_2$O$_3$-II type structure, \textit{Pbcn} orthorhombic) is observed \cite{ito_determination_2009,bykova_structural_2016}. 
Above $\sim$54 GPa, the $\iota$ phase transforms into $\zeta$-$\text{Fe}_2\text{O}_3$ phase (a distorted perovskite described in the monoclinic system \cite{bykova_novel_2013}) stable up to 55 GPa. Above 50-60 GPa, the $\zeta$-$\text{Fe}_2\text{O}_3$ phase transforms into $\eta$-$\text{Fe}_2\text{O}_3$ (\textit{Cmcm} post-perovskite orthorhombic), while above 67 GPa the possibly metastable $\theta$-$\text{Fe}_2\text{O}_3$ (\textit{Aba}2 orthorhombic), is also observed in a limited region between approximately 1000 and 2000~K~\cite{bykova_structural_2016}.

The behaviour of $\text{Fe}_2\text{O}_3$ under dynamic compression is less well understood and experimental results are scarce, largely due to the complexity of the target design. High-purity $\text{Fe}_2\text{O}_3$ samples in LH-DAC experiments are usually powders \cite{badro_nature_2002, sanson_local_2016, boulard_ferrous_2019, ono_situ_2005, ono_high-pressure_2004} or grown single-crystals \cite{bykova_structural_2016}.
In contrast, dynamic compression measurement require bulk, high-purity and homogeneous samples.
These are hard to find naturally with the required stoichiometry~\cite{kondo_electrical_1980}, but are also challenging to synthesize at the relatively large (mm) sizes needed in experiments~\cite{liebermann_elastic_1968}.
Nevertheless, gas gun measurements have been performed up to around 140~GPa \cite{mcqueen_handbook_1966, liebermann_elastic_1968} showing a significant volume drop of $\sim$10\% at approximately 50~GPa, which may be indicative of a phase transition. Another study also reported a significant drop in resistivity at $\sim$44-52~GPa using a double-stage light-gas gun on large natural $\text{Fe}_2\text{O}_3$ crystals~\cite{kondo_electrical_1980}.
However, the crystal structures of $\text{Fe}_2\text{O}_3$ under dynamic shock compression remain unknown, and it is unclear how these relate to the flurry of phases observed under static compression.
In this letter we present \textit{in-situ} time-resolved x-ray diffraction measurements in shock-compressed $\text{Fe}_2\text{O}_3$ using an x-ray Free Electron Laser (XFEL), at pressures between 38-122~GPa. Our results demonstrate that the behaviour of $\text{Fe}_2\text{O}_3$ is indeed very different under dynamic compression: we observe none of the high-pressure phases seen statically, and instead find a single new isostructural phase transition around 50-62~GPa, which we link to the collapse in magnetic ordering.

The experiment was performed at the Matter in Extreme Condition (MEC) end station of the Linac Coherent Light Source (LCLS)~\cite{nagler_matter_2015}. The shock was driven by two synchronized nanosecond lasers (527 nm) incident on the target at an angle of 20$\degree$ and with a spot size of 300~$\mu$m. We used pulse durations of 5, 10 and 15 ns, with a maximum total energy of 60~J on target. \textit{In-situ} x-ray diffraction (XRD) was performed using the x-ray beam operating in self-amplified spontaneous emission mode, with a photon energy of 7.08~keV, a spectral bandwidth of 25~eV, and pulse duration of 50~fs FWHM~\cite{madey_stimulated_1971,deacon_first_1977}. The x-rays were focused onto the target at an angle of 35$\degree$ with a spot size diameter of 60~$\mu$m, overlapping the focal spot of the optical lasers to probe the planar shock front region. The diffraction signal was measured in transmission geometry on 4 quadruple ePix10k detectors. Azimuthal integration of 2D diffraction images was carried out including solid angle correction, Al filter correction, polarization, and self-attenuation from the target. When this was not possible, integration was performed using the Dioptas software, based on the PyFai library~\citep{prescher_dioptas_2015,kieffer_new_2020}, including solid angle correction and polarization. Two Velocity Interferometer System for Any Reflector (VISAR) \cite{barker_laser_1972} were used to retrieve the velocity history and the time when the shock left the sample. VISAR sensitivities were 4.5241 and 1.9890 km s$^{-1}$ fringes$^{-1}$, respectively, with acquisition time windows of 10 and 20 ns.

Two different target designs were used. The first consisted of 8~$\mu$m of Fe$_2$O$_3$ deposited on 54~$\mu$m of parylen-N. The second contained a 8~$\mu$m layer of Fe$_2$O$_3$ sandwiched between a 22.1~$\mu$m sapphire window on one side, and a 54~$\mu$m layer of parylen-N on the other. A 200~nm aluminum flash coating was applied to the front face of all targets to minimize direct heating by the laser prepulse. 
The Fe$_2$O$_3$ was deposited via a physical vapor deposition process, leading to a polycrystalline structure of columnar crystallites with a preferential orientation along the $c$-axis of the crystal, and perpendicular to the shock wave propagation given the geometry of our setup (for further details see Figs.~S1-S8).

\begin{figure}[!h]
\includegraphics[scale=0.32]{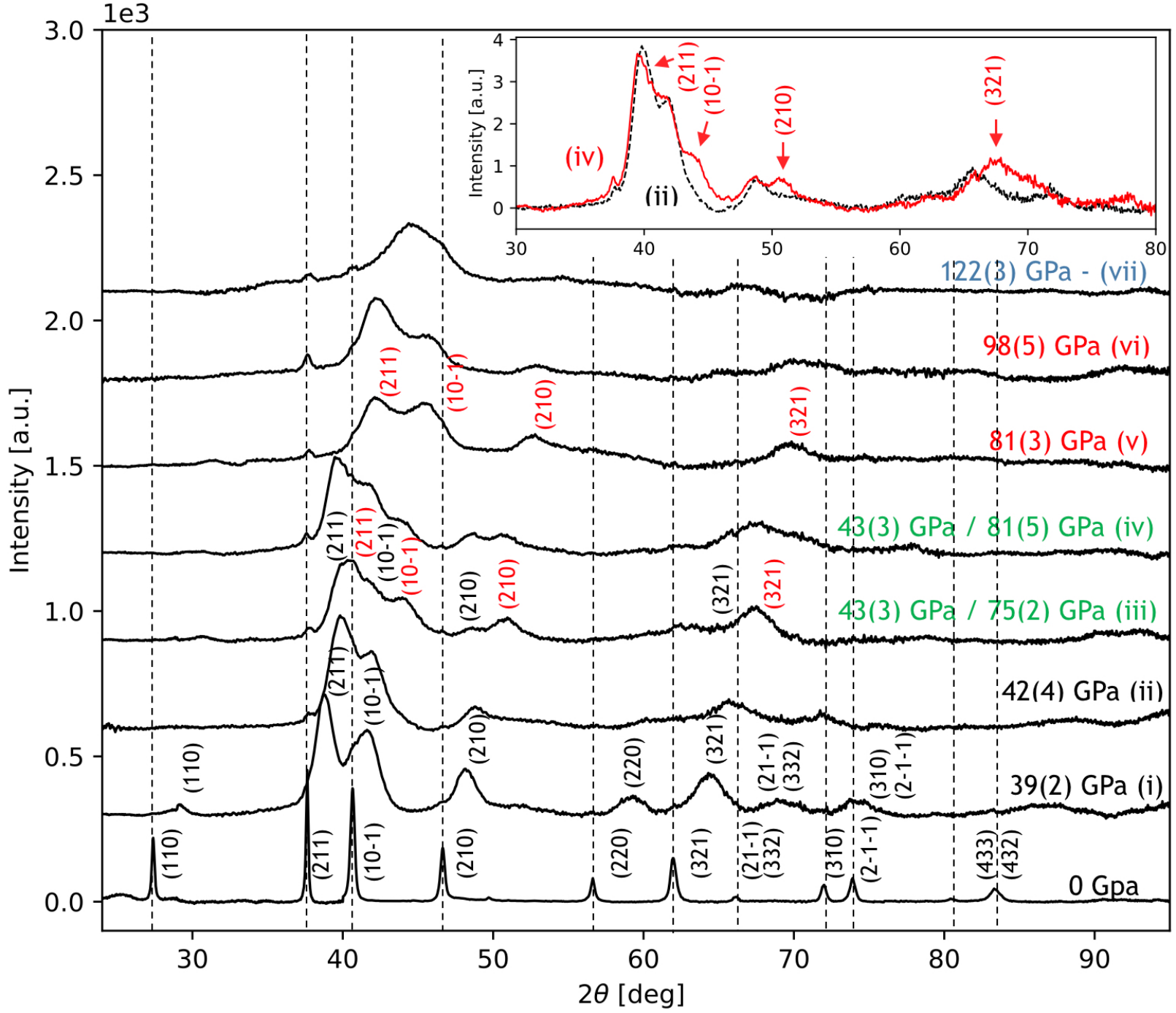}
\caption{\label{1d_Sachoc_lcls} Radially integrated x-ray diffraction profiles of $\text{Fe}_2\text{O}_3$ under shock for targets with a sapphire window between 0-122~GPa.
The vertical dotted lines corresponds to the ambient $\alpha$-$\text{Fe}_2\text{O}_3$ peak positions~\cite{blake_refinement_1966}. Miller (hkl) indices are displayed for the low- and high-pressure $\alpha$ structures in black and red, respectively. The inset highlights differences between patterns (ii) and (iv).}
\end{figure}

We show the radially integrated x-ray diffraction patterns of $\text{Fe}_2\text{O}_3$ under shock in Fig.~\ref{1d_Sachoc_lcls}, measured when nearly the entire $\text{Fe}_2\text{O}_3$ layer is shocked, but before shock breakout and release. The pattern recorded prior to the shock at ambient conditions is shown for comparison.
At pressures below 42~GPa, the patterns (i) and (ii) show similar features to that of the ambient sample in terms of both peak positions and intensities.
Nine $\alpha$-$\text{Fe}_2\text{O}_3$ peaks are identified and shifted toward larger angles due to sample compression. 
Overall, strong peak broadening is observed in shocked samples, reminiscent of a decrease in crystallite size and an increase in micro-strain effects (lattice parameters fluctuations) under pressure~\cite{ichiyanagi_microstructural_2019, briggs_ultrafast_2017}.
At intermediate pressures, patterns (iii) and (iv) show four additional peaks, indexed in red on pattern (iii). We attribute these to the appearance of a second kind of $\text{Fe}_2\text{O}_3$ phase, which we dub the $\alpha'$ phase, having the same crystallographic structure as $\alpha$ but a significantly lower volume. These mixed patterns are thus associated with two pressures depending on the phase.
For pressures above 81 GPa, only four peaks corresponding to the $\alpha^\prime$-$\text{Fe}_2\text{O}_3$ are observed, with a significant shift toward larger angles due to larger compression. At 122 GPa, pattern (vii) also shows the appearance of a diffuse signal at scattering angles of 45$\degree$ ($Q=2.7$~\AA$^{-1}$) and 55$\degree$ ($Q=3.4$~\AA$^{-1}$), which could be interpreted as an amorphous phase (2D diffraction data are shown in Fig.~S9).

\begin{figure}[h]
\includegraphics[scale=0.375]{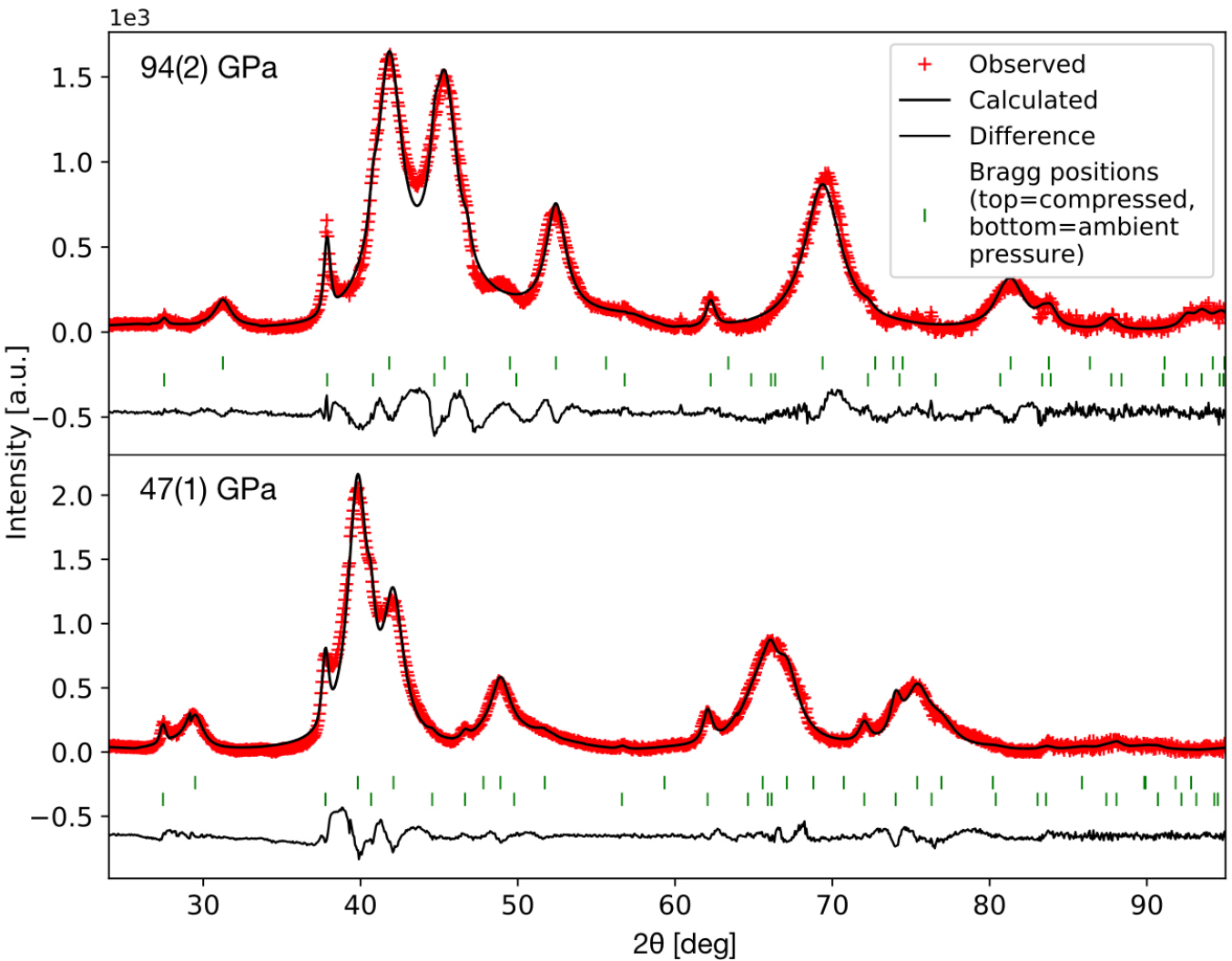}
\caption{Le Bail refinements of two radially integrated x-ray diffraction patterns just after shock breakout, at pressures of 47(1) GPa and 94(2) GPa.
Each pattern was fitted using two $\alpha$-$\text{Fe}_2\text{O}_3$ phases with $R\bar{3}c$ symmetry: one for the $\alpha$-$\text{Fe}_2\text{O}_3$ ambient at 0 GPa (bottom green tick marks) and one for the compressed $\alpha$-$\text{Fe}_2\text{O}_3$ phase (top green tick marks). 
%Red cross, thick black line, and thin black line represent the observed, calculated and difference patterns, respectively.
$R_\mathrm{p}$ factors were 10.5\% and 12.4\% for the 47 GPa and 94 GPa patterns, respectively.
%The best peak positions are shown as tick marks below each pattern with the ambient $\text{Fe}_2\text{O}_3$ at the bottom.
Fitted lattice parameters are $a_0=5.03$~Å; $c_0=13.03$~Å for ambient $\alpha$-$\text{Fe}_2\text{O}_3$; $a=4.87(2)$~Å; $c=12.94(2)$~Å for 47 GPa; and $a=4.56(2)$~Å; $c=12.54(2)$~Å for 94 GPa.}
\label{lebail}
\end{figure}

It was not possible to perform Rietveld refinements on the data due to the complex microstructure of the samples under shock. Instead, we performed Le Bail refinements, shown in Fig.~\ref{lebail}, for data from targets without a sapphire window.
At 47(1)~GPa a decrease of $c/a$ ratio for the compressed $\alpha$-$\text{Fe}_2\text{O}_3$ phase is observed compared with ambient $\alpha$-$\text{Fe}_2\text{O}_3$. At 94(2)~GPa we identify an $\alpha$-$\text{Fe}_2\text{O}_3$ structure with a significantly smaller volume than the $\alpha$-$\text{Fe}_2\text{O}_3$ observed at 47(1)~GPa (225.2~\AA$^3$ and 265.8~\AA$^3$, respectively).
In both cases, the $\alpha$-$\text{Fe}_2\text{O}_3$ structure~\cite{blake_refinement_1966} fits the compressed phase well ($R_\mathrm{p}=12.4$\%). For data above 80 GPa, none of the high pressure phases of $\text{Fe}_2\text{O}_3$ reported under static compression were observed ($\theta$-$\text{Fe}_2\text{O}_3$, $\eta$-$\text{Fe}_2\text{O}_3$, Fe$_{25}$O$_{32}$, or Fe$_5$O$_7$ \cite{bykova_structural_2016}).

\begin{figure}[h]
\includegraphics[scale=0.5]{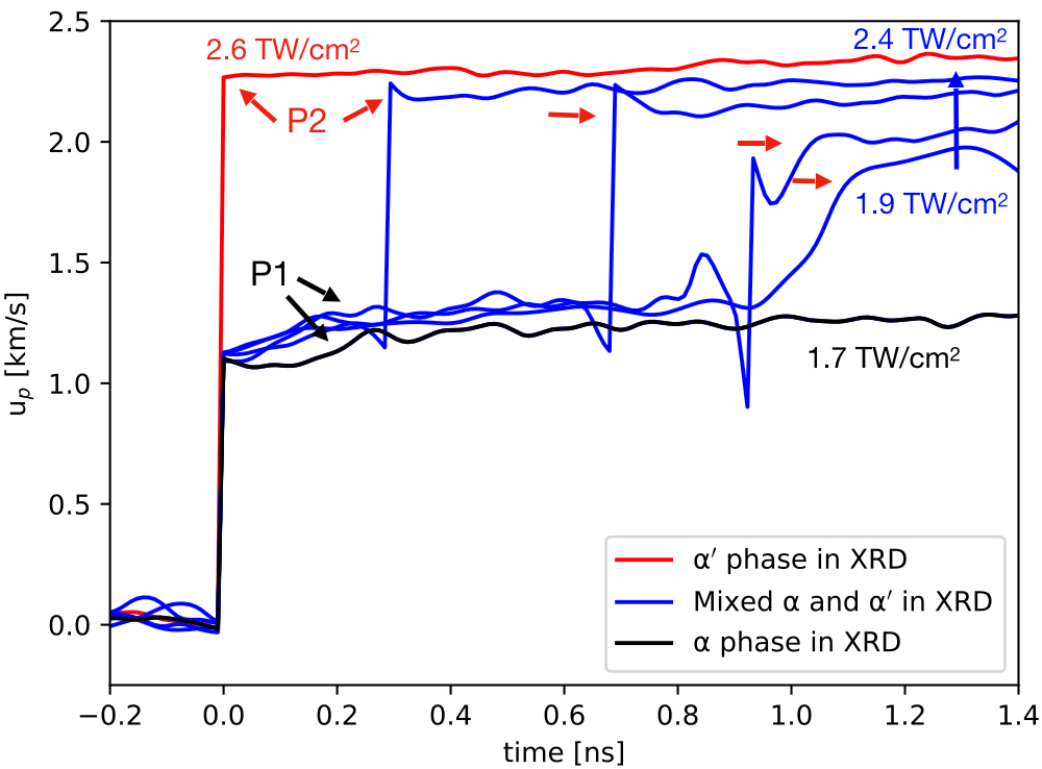}
\caption{\label{visar_phaseT} Particle velocities measured at the rear of the $\text{Fe}_2\text{O}_3$ sample for the different pressure regimes. The time origin is set to the shock breakout time in the sapphire.
In black we show a low-pressure shock where only a single $\alpha$ phase is observed in x-ray diffraction. The blue curves, taken at intensities between 1.9 and 2.4 TW/cm$^2$, correspond to thermodynamic conditions where both $\alpha$ and $\alpha'$ are observed in diffraction.
The red curve corresponds to conditions where only a single compressed $\alpha^\prime$ phase is observed. The double-wave structure is indicative of a phase transition.}
\end{figure}

Velocity profiles at the $\text{Fe}_2\text{O}_3$/Sapphire interface for increasing laser intensities are shown as a function of time in Fig.~\ref{visar_phaseT}.
A double-wave structure is systematically observed in VISAR data when the mixing of the $\alpha$ and $\alpha^\prime$ phases is detected in diffraction, as shown for shots in blue. The first wave is labelled P1 and the second wave P2. From the appearance of the $\alpha^\prime$ phase, the P1 wave velocity remains constant at around 1.3~km s$^{-1}$ for all mixed shots regardless of the laser intensity. In contrast, the P2 wave velocity increases with laser intensity and its arrival time becomes shorter.
The double-wave structure is also observed for targets without a sapphire window, measured directly from the free surface of $\text{Fe}_2\text{O}_3$ (Fig.~S8). This indicates that the double-wave structure is not due to the sapphire window, but is produced within the $\text{Fe}_2\text{O}_3$ layer at specific laser intensities.
The splitting of the wavefront into two waves is characteristic of a phase transition with a volume change \cite{zeldovich_shock_1967, mcqueen_equation_1970}.
We thus identify this observation as a signature of the $\alpha \rightarrow \alpha'$ phase transition, which is also observed in diffraction, and associate the P1 and P2 waves with the $\alpha$ and $\alpha^\prime$ phases, respectively.
The pressures in the $\alpha$ and $\alpha^\prime$ phases can then be determined from the P1 and P2 velocities, similarly to previous work on Bismuth~\cite{gorman_femtosecond_2018}.

\begin{figure}[h]
\includegraphics[scale=0.29]{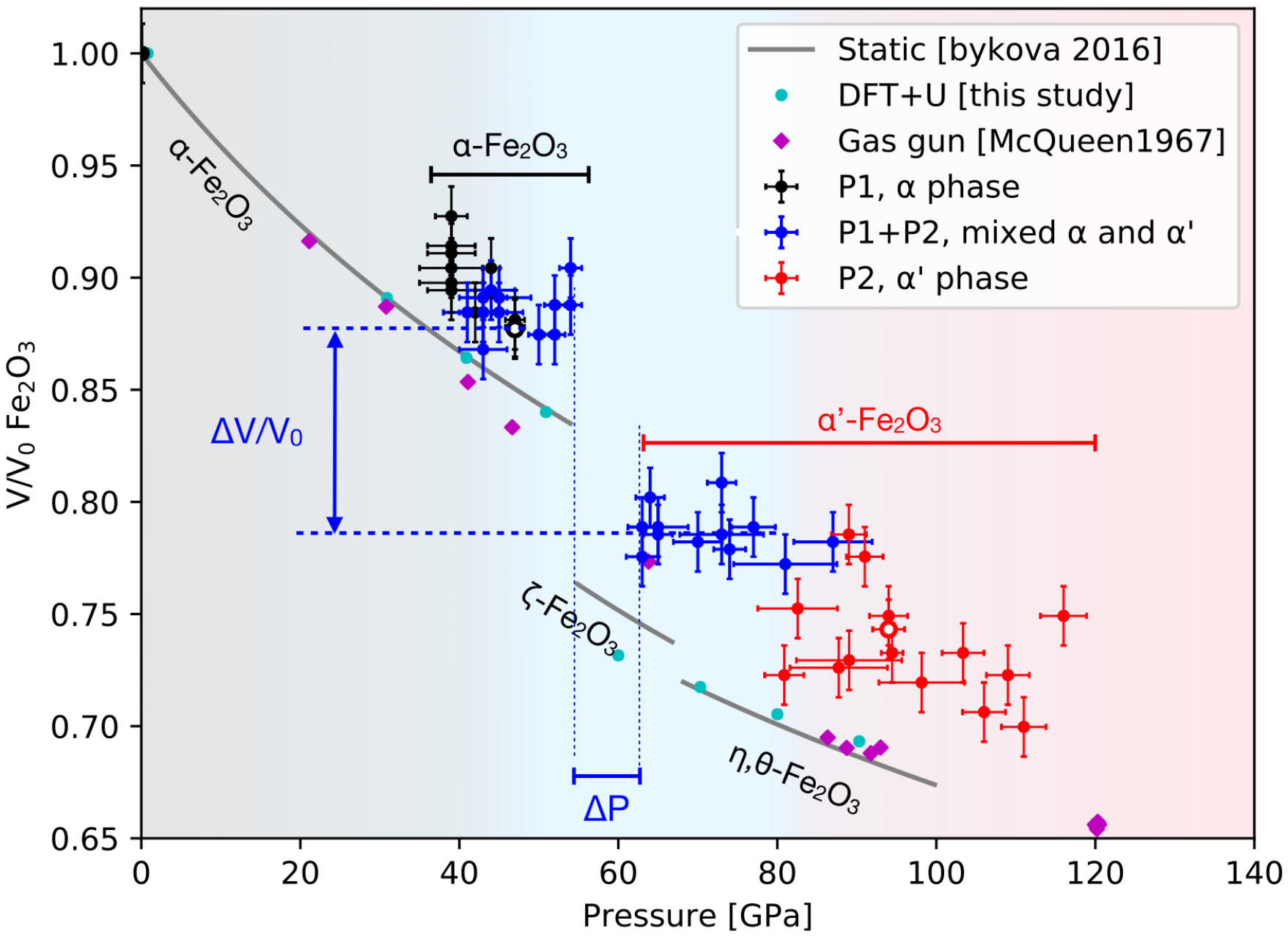}
\caption{\label{drx_phaseT} V/V$_0$ for $\alpha$-$\text{Fe}_2\text{O}_3$ and $\alpha^\prime$-$\text{Fe}_2\text{O}_3$ determined from x-ray diffraction as a function of pressure, compared with other published work. Points corresponding to the observation of only the $\alpha$-Fe$_2$O$_3$ phase are plotted in black, of only the $\alpha'$-Fe$_2$O$_3$ phase in red, and of the coexistence of both phases in blue. The two empty symbols correspond to data points determined from the Le Bail fits shown in Fig.~\ref{lebail}.
A transition from $\alpha$-$\text{Fe}_2\text{O}_3$ to $\alpha^\prime$-$\text{Fe}_2\text{O}_3$ phase is observed at 54-62~GPa. It is associated with a relative volume jump $\Delta V/V_0$ equal to 11.0~\% and pressure jump of $\Delta P\sim 9$~GPa.}
\end{figure}

The unit cell volumes of the $\alpha$-$\text{Fe}_2\text{O}_3$ and $\alpha^\prime$-$\text{Fe}_2\text{O}_3$ phases are plotted against pressure in Fig~\ref{drx_phaseT}. The $\alpha$-$\text{Fe}_2\text{O}_3$ phase is observed up to 54~$\pm$~2~GPa, and the $\alpha'$-$\text{Fe}_2\text{O}_3$ phase for pressures above 63~$\pm$~2~GPa.
We measure a (11.0~$\pm$~1.4)\% volume drop between the two phases, calculated from x-ray diffraction data showing both in coexistence. This volume drop indicates that the transition from $\alpha$ to $\alpha^\prime$ is a first-order transition.

The volume drop is accompanied by a pressure gap of around 9~$\pm$~3~GPa.
%Although the $\alpha$ and $\alpha^\prime$ phases are both seen in x-ray diffraction, no P2 wave associated with such coexistence could be detected in the VISAR in this pressure gap. 
This could be due to experimental detection limitations, such as 1) the VISAR etalon; 2) the temporal range restricted by the multiple waves in our sample design (see Fig.~S4 of the supplementary) that might prevent the detection of the P2 waves with breakout time superior to the temporal range; or 3) the inherent laser-shocked x-ray diffraction peak broadness that prevents small $\alpha^\prime$ peaks to be observed. In addition, phase transitions under shock are also subject to temperature effects that might broaden the transition pressure range~\cite{zeldovich_shock_1967}, as well as superheating, kinetic and plasticity effects~\cite{pepin_kinetics_2019, mcbride_phase_2019, smith_time-dependence_2013} that might affect the onset pressure for $\alpha^\prime$-$\text{Fe}_2\text{O}_3$ synthesis.

Overall, our data yield larger unit cell volumes than previous shock data from gas gun experiments~\cite{mcqueen_handbook_1966} as seen in Fig.~\ref{drx_phaseT}. However, these previous measurements are intrinsically different to ours, either due to their time scale (ms or $\mu$s for gas guns) or by the nature of the measurements (optical and macroscopic compared with x-ray bulk and microscopic). In this work, the volume is extracted from the positions of the diffraction peaks. Further, we note that $\text{Fe}_2\text{O}_3$ single-crystals are known to behave differently depending of the relative alignment of the compression direction with the crystal axis~\cite{kondo_electrical_1980}. Because of how our samples were deposited they do have a notable preferred orientation, and this may well affect the experimental compression values.
% Note sure what is the value of the sentence below – it doesn't really say much, perhaps we can cut it.
%Our data are also shifted in compared to static data. If no particular agreement is expected above 50 GPa as phases are different, the disagreement from 38 to 50 GPa where $\alpha$-$\text{Fe}_2\text{O}_3$ phase is observed both in static and in dynamic compression can be explained by the same phenomena stated above. 

\begin{figure}[h]
\includegraphics[scale=0.30]{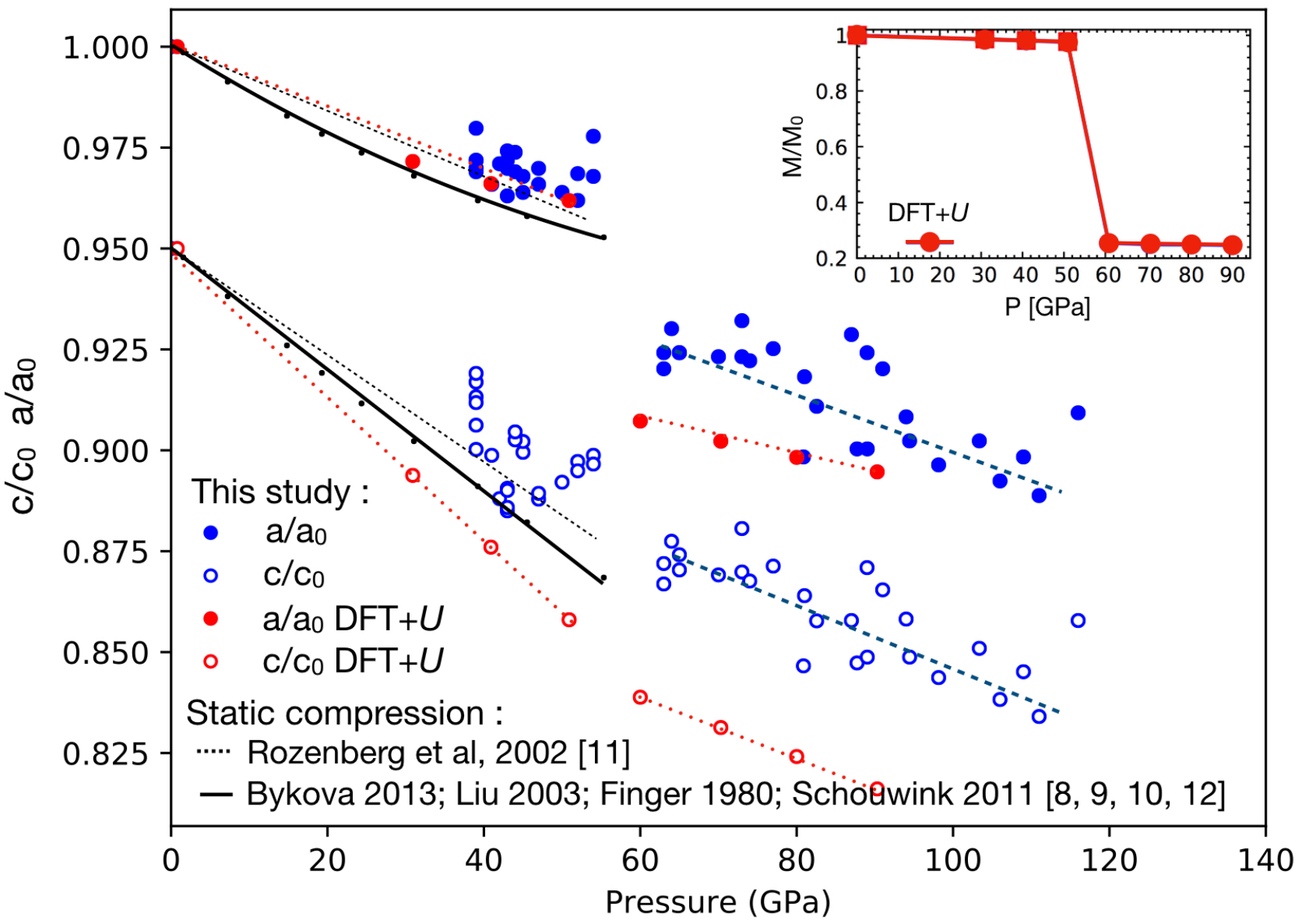}
\caption{\label{drx_a_c} Lattice parameters for the $\alpha$ and $\alpha'$ phases as a function of pressure. All c/c$_0$ data are shifted vertically by -0.05 for clarity. Shock data are represented by solid blue dots (a/a$_0$) and empty blue dots (c/c$_0$). Dashed blue lines above 60~GPa are drawn to guide the eye.
Lattice parameters for the $\alpha$ phase from static compression are given as solid black lines for ref.~\cite{bykova_novel_2013, liu_static_2003, schouwink_high-pressure_2011, finger_crystal_1980} and black dotted line for ref.~\cite{rozenberg_high-pressure_2002}. The results from DFT$+U$ using a rhombohedral phase are given in red. The inset figure shows the relative magnetic moment of Fe from DFT$+U$ as a function of pressure. The phase is high spin below 50~GPa, and low spin above 60~GPa. }
\end{figure}

Figure~\ref{drx_a_c} shows the evolution of $a/a_0$ and $c/c_0$, where $a$ and $c$ denote the lattice parameters under shock and $a_0$ and $c_0$ the parameters at ambient conditions. Both are deduced from Le Bail refinements. Below 54 GPa, the ratios for $\alpha$-$\text{Fe}_2\text{O}_3$ are consistent with static compression data~\cite{rozenberg_high-pressure_2002}, and can be explained by bonding distortion or uneven modifications of the Fe-O-Fe bond lengths~\cite{bykova_novel_2013, liu_static_2003, schouwink_high-pressure_2011, rozenberg_high-pressure_2002, finger_crystal_1980}. This pressure distortion effect is observed in other corundum-structured oxides~\cite{bykova_novel_2013, cox_transition_2010}, and can lead to a Mott insulator transition.
A discontinuity of 3.6~\% is seen between 54-62~GPa for a/a$_0$, in line with the observed volume drop between the $\alpha$ and $\alpha^\prime$ phases shown in Fig.~\ref{drx_phaseT}. No discontinuity is observed for c/c$_0$. Above 62 GPa we observe a change in the slope of c/c$_0$ with pressure, while the slope of the corresponding a/a$_0$ line remains unchanged. This indicates that the {\it c}-axis becomes less compressible in the new $\alpha^\prime$ phase, and that the volume drop observed is mostly due to a strong decrease of the \textit{a} lattice parameter.

As the transition is observed without change of symmetry, the $\alpha$-$\text{Fe}_2\text{O}_3$ to $\alpha^\prime$-$\text{Fe}_2\text{O}_3$ transition could be electronically driven. Two electronic transitions are observed at 50~GPa in static compression experiments: a Mott transition and a simultaneous spin transition~\cite{greenberg_pressure-induced_2018}. These transitions are not limited by the timescale required for atomic displacement, and can therefore be much faster than a structural transition. Moreover, it has been shown that Mott and spin transitions can lead to a significant volume collapse without structural change~\cite{greenberg_pressure-induced_2018}, as seen for example in MnO~\cite{yoo_first-order_2005} and FeO~\cite{ohta_experimental_2012}.
To further investigate such electronic transitions, we performed DFT$+U$ calculations~\cite{Hohenberg,Kohn,Anisimov,Liechtenstein,Dudarev,Cococcioni} of $\alpha$-$\text{Fe}_2\text{O}_3$. As our experimental data show only $\alpha$-$\text{Fe}_2\text{O}_3$ phase, relaxation of cell parameters, atomic positions and magnetic state of iron was performed with $R\bar{3}c$ symmetry. Our simulations were performed using the v7.1 Quantum-Espresso suite~\cite{QE,QE2}. Scalar-Relativistic ultrasoft pseudopotential (PP) \cite{UPF} were generated using the PBEsol exchange-correlation functionals \cite{PBEsol} for Fe and O. The effective Hubbard $U$ parameter was used for the Fe-$3d$ orbitals, with the initial occupations given by the PP. We used a kinetic-energy cutoff of the plane-wave basis set of 100~Ry and an augmentation charge energy cutoff of 800~Ry. The calculations were carried out using a 12$\times$12$\times$12 $\boldsymbol{k}$-point grid. 
The DFT+$U$ calculations were performed at ambient pressure using $U=5$~eV, leading to an energy band gap of 2.075~eV and magnetic moment per Fe atom of 4.41~$\mu_\mathrm{B}$, in agreement with experimental values of 2.14~eV and 4.6~$\mu_\mathrm{b}$, respectively \cite{Benjelloun,Coey}.

The results of the DFT$+U$ calculations are shown in Figs.~\ref{drx_phaseT} and \ref{drx_a_c}. All main features of interest observed experimentally around the phase transition at 50-60~GPa are reproduced in the simulations: we find a large V/V$_0$ volume drop ($\sim$10\%), a discontinuity in the evolution of a/a$_0$ but without a change in compressibility, and a change in compressibility but without a discontinuity of c/c$_0$.
While there remains some discrepancy in the absolute values for the calculated parameters, overall the simulations reproduce the observed trends with good fidelity. The volume drop between 50-60~GPa is correlated to a drop in the local magnetic momentum of iron, as shown in the inset in Fig.~\ref{drx_a_c}, decreasing by a factor of 5, i.e. changing from a high to a low spin state. Therefore, we posit that the 
$\alpha \rightarrow \alpha'$ transition could be explained by a spin transition from a high-spin rhombohedral phase ($\alpha$-$\text{Fe}_2\text{O}_3$) to a low-spin rhombohedral phase ($\alpha^\prime$-$\text{Fe}_2\text{O}_3$ being a low-spin corundum-structured phase).
We note that, in principle, the value of U in our simulations should change with pressure. We performed similar calculations for different values of U (1,3,5,7~eV), and find that while this affects the pressure at which the transition takes place, it does not change the calculated volume drop of $\sim$10\%, nor the change from high to low spin. While our DFT$+U$ calculations are not tailored to identify a Mott transition, it is known from dynamical mean-field theory calculations that the spin transition and the Mott transition are linked and occur simultaneously~\cite{kunes_pressure-driven_2009}. 
We thus propose that the $\alpha \rightarrow \alpha'$ transition is an electronically driven spin transition from high-spin to low-spin, possibly associated with a Mott transition. Importantly, this implies that the electronic transition will occur before any structural transition takes place.
Static compression experiments reported simultaneous electronic (spin and Mott transitions) and structural transitions (to $\eta$ or $\theta$-$\text{Fe}_2\text{O}_3$ phases)~\cite{bykova_structural_2016, ono_situ_2005, ono_high-pressure_2004, pasternak_breakdown_1999, badro_nature_2002, sanson_local_2016, greenberg_pressure-induced_2018}. Here our time-resolved experiments show an electronic transition occurring before such structural transitions, proving that the electronic transition implies the structural change as indicated by~\cite{sanson_local_2016}, and not the other way around.

To summarize, we have presented \textit{in-situ} time-resolved x-ray diffraction measurements in shock-compressed $\text{Fe}_2\text{O}_3$ up to 122~GPa, showing a clear difference in the phase diagram compared with static compression experiments. None of the high pressure phases seen statically where observed dynamically. Instead, we observed an isostructural $\alpha \rightarrow \alpha'$ phase transition around 50-60~GPa, characterized by an 11\% volume drop due to a high-spin to low-spin transition, and, possibly, a Mott transition. Our results thus show that the electronic transition(s) observed in $\text{Fe}_2\text{O}_3$ under static compression around 50~GPa still occur, but without the associated structural transitions.
These either cannot happen under dynamic laser compression, or require timescales longer than those accessible in our experiment to form. The structures of the $\theta$ and $\eta$ phases differ significantly from the structure of the $\alpha$ phase: they are composed of FeO$_6$ prisms and octahedra, whereas the $\alpha$ phase is only composed of FeO$_6$ octahedra~\citep{bykova_structural_2016}. Our results thus suggest that the fast electronic transitions ($<$1~ns) are the ones driving the comparatively slower reconstructive structural transitions to the $\theta$ and $\eta$ phases ($>$1~ns)~\citep{burns_space_2013}, and not vice versa.

Use of the Linac Coherent Light Source (LCLS), SLAC National Accelerator Laboratory, is supported by the U.S. Department of Energy, Office of Science, Office of Basic Energy Sciences under Contract No. DE-AC02-76SF00515. 
This project has received funding from the European Research Council (ERC) under the European Union’s Horizon 2020 research and innovation program (ERC PLANETDIVE grant agreement No 670787). This work was supported by grants from Japan Society for the Promotion of Science (JSPS).
C.C., S.A., P.H., J.S.W and S.M.V. acknowledge support from the UK EPSRC under grants EP/P015794/1 and EP/W010097/1.
T.G. acknowledges support from AWE via the Oxford Centre for High Energy Density Science (OxCHEDS).
T.C. and S.M.V. acknowledge support from the Royal Society.
A.F. acknowledges support from the STFC UK Hub for the Physical Sciences on XFELs.
We thank the microscopy, x-ray diffraction and PVD platforms at IMPMC for support in producing and characterizing the $\text{Fe}_2\text{O}_3$ samples. We also thank T.~De Resseguier, R. Smith,  T.~Vinci and A.~Benuzzi-Mounaix for helpful discussions.

\bibliography{refs_lcls_paper}% Produces the bibliography via BibTeX.
\end{document}